\documentclass[aps,prd,preprint,preprintnumbers,showpacs,showkeys,nofootinbib,superscriptaddress]{revtex4-1}
\usepackage{multirow}
\usepackage{booktabs}    
\usepackage{placeins}
\usepackage{soul}
\usepackage[usenames,dvipsnames,svgnames]{xcolor}
\usepackage[colorlinks]{hyperref}
\usepackage{bbm}
\usepackage{amsfonts}
\usepackage{amsmath}
\usepackage{mathrsfs}
\usepackage{soul}

\usepackage{amsmath,amssymb,amsbsy,bm}
\usepackage{graphicx}

\newcommand{\diag}{\mathop{\mathrm{diag}}}

\begin{document}

\title{\boldmath A Tale of Two-U(1) Axion Models} 
\author{Dong Hu}
\email{hudong16@pku.edu.cn}
\affiliation{Department of Physics and State Key Laboratory of Nuclear Physics and Technology, Peking University, Beijing 100871, China}

\author{Hao-Ran Jiang}
\email{h.r.jiang@pku.edu.cn}
\affiliation{Department of Physics and State Key Laboratory of Nuclear Physics and Technology, Peking University, Beijing 100871, China}

\author{Hao-Lin Li}
\email{lihaolin@itp.ac.cn}
\affiliation{CAS Key Laboratory of Theoretical Physics, Institute of Theoretical Physics, Chinese Academy of Sciences, Beijing 100190, P. R. China}

\author{Ming-Lei Xiao}
\email{mingleix@itp.ac.cn}
\affiliation{CAS Key Laboratory of Theoretical Physics, Institute of Theoretical Physics, Chinese Academy of Sciences, Beijing 100190, P. R. China}

\author{Jiang-Hao Yu}
\email{jhyu@itp.ac.cn}
\affiliation{CAS Key Laboratory of Theoretical Physics, Institute of Theoretical Physics, Chinese Academy of Sciences, Beijing 100190, P. R. China}
\affiliation{School of Physical Sciences, University of Chinese Academy of Sciences, Beijing 100049, P.R. China}
\affiliation{School of Fundamental Physics and Mathematical Sciences, Hangzhou Institute for Advanced Study,  University of Chinese Academy of Sciences, Hangzhou 310024, China} 
\affiliation{International Center for Theoretical Physics Asia-Pacific, Beijing/Hanzhou, China}

\begin{abstract}

The $U(1)$ global symmetry to solve the strong $CP$ problem could be the remnant of multi-$U(1)$ symmetries from QCD and hidden strong dynamics. Both Peccei-Quinn $U(1)$ and dynamical $U(1)$ are described uniformly, based on which we classify various mixed two-$U(1)$ models to solve both strong $CP$ and quality problems. We propose a moose diagram method with different fermion assignments to directly read relations between $CP$ phases, which illustrate how strong $CP$ problem is solved in terms of cancellation between $CP$ phases. In two axion models, we find the lightest axion is still same as QCD axion at infrared region, while one axion model with $Z_2$ symmetry enhances the axion mass spectrum. Our discussions can be extended to multi-axion cases.

\end{abstract}

\maketitle
	\section{Introduction}
	In the 1970s, 't Hooft proposed that QCD has a nontrivial vacuum structure and solved the $U(1)_A$ problem~\cite{Hooft1976,Hooft1976a}. The nontrivial vacuum structure suggests that there is an additional topological term which violates $CP$ symmetry in the QCD Lagrangian. The $CP$ violating term brings a free parameter denoted as $\theta$. Due to the axial anomaly, a chiral rotation of quarks changes $\theta$ and the argument of Yukawa couplings of quarks, while the sum of them is invariant. Thus, the observable strong $CP$ phase in the Standard Model (SM) is
	\begin{equation} \label{eq: SM_theta}
	\overline{\theta}=\theta+\arg \det (Y_{u} Y_{d}) ,
	\end{equation}
	where $u,d$ represent up and down type quarks respectively. The measurement of neutron electric dipole moment (EDM) suggests that $|\bar\theta| < 10^{-10}$~\cite{Baker2006}. How to understand the extreme smallness of $\overline{\theta}$ is the well-known strong $CP$ problem.

	There are several ways to solve the strong $CP$ problem. One solution is the Nelson-Barr mechanism~\cite{Nelson1984NWCV,Barr1984StSwtPQS}, which assumes $CP$ to be conserved at high energy scale, i.e. the $CP$-violating phases of the CKM matrix and $\overline{\theta}$ are both zero at the high energy scale. While at the low energy scale the $CP$-violating phase $\delta$ of the CKM matrix is reproduced and $\overline{\theta}$ is still fixed to be zero. 
	An alternative way is utilizing the chiral rotation of fermions to absorb $\overline{\theta}$, when a massless quark or an additional global chiral $U(1)$ symmetry exists. However, the solution with massless quark has been disfavored by the results of Lattice QCD with $m_u= 2.2 ^{+0.5}_{-0.4}~MeV$~\cite{Gasser1982}. The additional global chiral $U(1)$ solution was first proposed by Peccei and Quinn~\cite{Peccei1977,Peccei1977a}. Later, Weinberg~\cite{Weinberg1978ANLB} and Wilczek~\cite{Wilczek1978PoSitPoI} predicted a pseudoscalar as the Goldstone boson called axion from the spontaneous breaking of this additional $U(1)_{PQ}$ symmetry. Although the original Weinberg-Wilczek axion model has been ruled out by experiments~\cite{Bardeen1987}, derived models still survive. The KSVZ~\cite{Kim1979a,Shifman1980a} and the DFSZ~\cite{Dine1981a,zhitnitskii1980possible} are the most typical models among them.
	
	Although the KSVZ and the DFSZ survive from experimental constrains, there exists an additional theoretical issue, the quality problem. Constraints from Astrophysics give the lower bound of the decay constant of axion to be $f_{a}\gtrsim  10^{8} \mathrm{GeV}$~\cite{RaffeltAAB}. There is a general consensus that gravitational effects generate operators suppressed by the Planck scale $M_{\mathrm{Pl}}$, which  explicitly break global symmetries. As for the axion model, the explicit breaking effect is estimated by the operators~\cite{Barr1992,Kamionkowski1992}
	\begin{equation}\label{eq: gravitational violate}
	V(\phi)=g\frac{|\phi|^{2 m} \phi^{n}}{M_{\mathrm{Pl}}^{2 m+n-4}}+h . c . .
	\end{equation}
	Although these operators are suppressed by the Planck scale, the smallness of strong $CP$ phase and high scale of $\langle\phi\rangle\sim f_{a}$ cause non-negligible effects from the lower dimension operators. Consequentially, the minimum of the scalar potential is shifted and the $CP$ phase $\overline{\theta}$ reappears.
	In order to solve this problem, one can impose a discrete symmetry $Z_N$ with large $N$ on $\phi$ so that operators with dimension less than $N$ are forbidden. For $f_a=10^{12}$ $\mathrm{GeV}$, a scale that the axion can serve as dark matter, operators with dimension $2m+n < 14$ are forbidden~\cite{Hook2018TLotSCPaA}. Another idea is to reduce the VEV $\langle\phi\rangle$ while preserving a large $f_a$. The implementation is accomplished in the multiple axion models using the alignment mechanism ~\cite{Kim2005,Choi2014a,Higaki2016} or the Clockwork mechanism~\cite{Bonnefoy2018}. On the other hand, very heavy aixon with $m_{a} \gtrsim O(100) \mathrm{MeV}$ evades the astrophysical constraints~\cite{Fukuda2015AMoVQA}, thus a small $f_a$ is allowed to relax the quality problem. Furthermore, in some composite axion models, gauge invariance forbids operators of high dimension by arranging fermions suitably ~\cite{Randall1992CAMaPSP}. In this paper we focus on the simplest multiple axion model, two axion model.
	
	On the other hand, additional non-Abelian gauge group is often introduced in various new physics, such as Hidden Valley~\cite{Strassler2006EoaHVaHC}, Vectorlike  Confinement~\cite{Kilic2009VCatL} and Twin Higgs~\cite{Chacko2006NEBfaMS}. These extensions of the SM are usually motivated by the Hierarchy problem or dark matter. Furthermore, the $U(1)$ global symmetry to solve the strong $CP$ problem could be the remnant of multi-$U(1)$ symmetry from the hidden strong dynamics. We call this kind of non-Abelian gauge groups as ``hidden QCD" in this paper.


	Similar to the SM QCD, the hidden QCD may also contain the $CP$ violation source from the $\theta'G'\tilde{G'}$ term and the relevant fermion Yukawa couplings, though it may or may not influcence the observed neutron EDM. Besides the QCD theta term, there are three types of effective operators directly contributing to the neutron EDM, including the quark EDM, the quark chromo EDM and the three-gluon Weinberg operator~\cite{Weinberg1989LHBETitNEDM,Pospelov2005EDMAPoNP}:
	\begin{equation}\begin{aligned}
	O_d&=-\frac{i}{2}  d  \bar{q} \sigma^{\mu \nu} \gamma_{5} q F_{\mu \nu}, \\
	O_{\tilde{d}}&=-\frac{i}{2}  \tilde{d}  \bar{q} \sigma^{\mu \nu} t^{a} \gamma_{5} q G_{\mu \nu}^{a}, \\
	O_w&=\frac{1}{3} w f^{a b c} G_{\mu \nu}^{a} \widetilde{G}_{\nu \beta}^{b} G_{\beta \mu}^{c},
	\end{aligned}\end{equation}
	where $d$ denotes EDM, $\tilde{d}$ denotes chromo EDM, $t^a$ is the generator of the QCD group and $f^{abc}$ denotes the QCD structure constant. The $\theta'G'\tilde{G'}$ term in hidden QCD is not directly related to these operators. However, two typical scenarios will result in the sensitivity of the neutorn EDM to the CP violation in the hidden sector:
	\begin{itemize}
		\item  Two strong $CP$ angels can be associated with each other by introducing a pseudo-scalar. The interaction between the pseudo-scalar and gauged fermions and the chiral rotation of the pseudo-scalar are 
		\begin{equation}\label{first way}
		\begin{aligned}
		\mathcal{L}&\sim y_q e^{ina/f_a} q\bar{q} + y_Q e^{ima/f_a} Q\bar{Q}+h.c. , \\
		a/f_a&\rightarrow a/f_a+\alpha,
		\qquad \theta \rightarrow \theta-n\alpha,
		\qquad \theta' \rightarrow \theta'-m\alpha,
		\end{aligned}
		\end{equation}
		where $q$ and $Q$ notate fermions charged under QCD and hidden QCD respectively. $a$ is the pseudo-scalar. $n$ and $m$ are constants.
		\item Fermions charged under both QCD and hidden QCD can link these two $CP$ phases. The chiral rotation of these fermions can change both two phases as
		\begin{equation}\label{second way}
		\psi_{L}(N_c,N_h)\rightarrow \psi_{L}(N_c,N_h) e^{i\beta} , \qquad \theta \rightarrow \theta-N_h\beta ,
		\qquad \theta' \rightarrow \theta'-N_c\beta,
		\end{equation}
		where $N_c$ and $N_h$ are representation of fermions in QCD and hidden QCD respectively.
	\end{itemize}
	Under such circumstance,  both $\theta $ and $\theta ^\prime $ have observable impact on the strong $CP$ problem, as either the Peccei-Quinn symmetry or the chiral rotation of massless fermion can transfer the $\theta (\theta')$ term into each other. Consequently, the physical $CP$-violating angle in QCD related to the neutron EDM is the linear combination
of the two theta angles. Therefore,
 these two angles are needed to be small simultaneously to explain the strong $CP$ problem. 
Either two global chiral $U(1)$ symmetries are needed, or there is one chiral $U(1)$ symmetry with a $Z_2$ among the two sectors. In this work, we will focus on the first scenario, while the later one is discussed in Ref.~\cite{Rubakov1997,Berezhiani2001a,Albaid2015SCaS,Fukuda2015AMoVQA,Hook2014ASttSCP,Chiang2016}.

	Chiral $U(1)$ symmetries are classified into two kinds, according to whether the corresponding axion is elementary or composite. 
	We refer to the one generating the elementary axion as the $U(1)_{PQ}$ arinsing from the phase of a complex scalar and the one generating composite axion as (also called dynamical axion ~\cite{Choi1985DA}) the $U(1)_A$ arising when massless fermions of the hidden QCD condensate at the high energy scale.
	In infrared region, both scenarios share similar nature and the pseudoscalars are the Goldstones of these chiral $U(1)$ symmetries named axion. Instanton effects explicitly break these symmetries and determine the properties of axion such as mass and axion-photon coupling. Therefore, new instanton effects of the hidden QCD could enlarge axion mass in some models with $Z_2$ symmetry~\cite{Rubakov1997,Fukuda2015AMoVQA,Gherghetta2016a} 
	and  it could also enhance axion-photon coupling in~\cite{Agrawal2017ETfPCotQA}. Moreover, an additional strong dynamics also provides solutions to quality problem and domain wall problem~\cite{Sikivie1982ADWatEU}.
	

	With hidden QCD introduced, we find new mixed $U(1)$ solutions which contain two sorts of chiral $U(1)$ symmetries. In these cases, spontaneous symmetry breaking of $U(1)_{PQ}$ could be triggered by the dynamical symmetry breaking in the hidden sector, besides the conventional Ginzburg-Landau potential method. The cancellation of $CP$ phases is always viable, except alignment situation. The cancellation of $CP$ phases in the QCD and the hidden QCD sector is shown in section~\ref{section3}. The lightest axion in these models are similar to the QCD axion. Some phenomena of these models are discussed in section~\ref{section4}.

	In addition to the new mixed $U(1)$ solution mentinoed above, we also propose a ``moose-like'' diagram method to visualize the cumbersome relations among gauge groups, new fields and CP phases in multi-axion models, from which the $U(1)$ charges for the fields and the potential of the relevant axions and CP phases can be easily read. Moreover, the diagram method helps to construct models containing more $U(1)$ and gauge groups.


	This paper is organized as follows. In section~\ref{section2}, we discuss two typical patterns to realize Chiral $U(1)$. Next, in section~\ref{section3}, we give solutions of the strong $CP$ problem, and propose a diagram method to present these solutions.
	In section~\ref{section4}, we study the axion mass, axion-photon coupling and axion deacy constant of one-axion solutions and two typical mixed two-axion solutions.
	Finally, we summarize the result in section~\ref{section5}.
	
	\section{Pecci-Quinn or Dynamical Solution}
	\label{section2}
	A large class of models solve the strong $CP$ problem by adding additional global $U(1)$ symmetries. There are mainly two kinds of $U(1)$s: the one in the Pecci-Quinn mechanism~\cite{Peccei1977,Peccei1977a}, denoted as $U(1)_{PQ}$, is associated with a elementary scalar $\phi$; the other is the axial $U(1)_A$ of some massless fermions charged under gauge groups, which induces the dynamical solution~\cite{Choi1985DA}. 
	
	Although $U(1)_{PQ}$ and $U(1)_A$ symmetries have different origins in ultraviolet region, they solve the strong $CP$ problem with the same philosophy: to make the $CP$ phase dynamically cancelled by introducing a $U(1)$ pseudo-Goldstone boson (PGB) with an anomaly-induced potential. For both scenarios, there are fermions axially charged under the global $U(1)$, 
	\begin{equation}\label{anormalous transformation}
	 f_L \rightarrow e^{i\alpha}f_L, \qquad f_R \rightarrow e^{-i\alpha}f_R.
\end{equation}
	However, the $U(1)$ is broken in different ways: for Pecci-Quinn mechanism, it is broken by the scalar potential $V(\phi)$, and the PGB turns out to be $a \sim\arg\phi$, the axion; in the dynamical solution, it is broken by condensation of the fermions, and the PGB is a composite of the fermion, also known as the dynamical axion. The common infrared behavior is that both PGB correspond to some axial currents, whose conservation is broken only by anomaly. It inspires a general description of these models, as presented in the following.
	
\begin{figure}[h]
		\centering
		\includegraphics[width=1\linewidth]{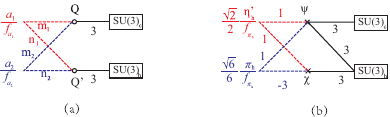}
		\caption{ In diagram (a), the solid line between Q and $SU(3)_c$ shows that Q is under representation 3 of QCD and the solid line between $Q^\prime$ and $SU(3)_h$ shows that Q is under representation 3 of hidden QCD. The red dashed line between $a_1$ and Q($Q^\prime$) shows that the charge of Q($Q^\prime$) under $U(1)_{PQ}$ is $m_1$($n_1$). Similarly, blue dashed line shows that the charge of Q($Q^\prime$) under $U(1)_{PQ^\prime}$ is $m_2$($n_2$). $1/f_{a_1}$ and $1/f_{a_2}$ are scale factors for two $U(1)$s. In diagram (b), solid lines show that $\psi$ is under representation 3 for both QCD and hidden QCD, and $\chi$ is a singlet for QCD but under representation 3 for hidden QCD. Red dashed lines show that the ratio of $\psi$ and $\chi$ in $\eta_h^\prime$ current is 1:1, which is similar to $U(1)$ charge in the left one at infrared region. The only difference is one more normalization factor $\sqrt{2}/2$ in front of the scale factor $1 / f_{\pi_h}$. Similarly, blue lines show that the ratio of $\psi$ and $\chi$ in $\pi_h$ current is 1:-3. The normalization factor is $\sqrt{6}/6$ and the scale factor is $1 / f_{\pi_h}$.} 
		\label{fig:figure1}
	\end{figure}

	In this section, we propose a ``moose-like'' diagram method to uniformly illustrate the structures of the models solving the strong $CP$ problem. The notation of all possible additional fermions introduced are listed in TABLE~\ref{tab: fermion classify}. As an example, we show the structure of the aligned axion model(left)~\cite{Kim2005} and the dynamical axion model(right)~\cite{Choi1985DA} in Fig.~\ref{fig:figure1}.  At the left end, we list all the axions, each representing a global $U(1)$, normalized as their proper contributions to the CP phase. Dashed lines link the axions to the fermions, and the numbers on  them indicate the corresponding $U(1)$ charges of the fermions. The solid lines with tags show the non-trivial representations of the fermions under the linked gauge groups at the right end, like the color $SU(3)_c$ and the hidden $SU(3)_h$. Circles ($\circ$) on the vertices represent massive fermions; Crosses ($\times$) on the vertexes represent massless fermions. 
	
	\begin{table}[htbp]
		\caption{Fermion charge and representation}
		\centering
		\begin{tabular}{c|ccc}
			\hline
			& {$SU(3)_c$} & {$SU(3)_h$} & $U(1)_{PQ}$ \\
			\hline
			\hline
			$\psi$ & 3     & 3     & 0 \\
			$\chi$ & 1     & 3     & 0 \\
			\hline
			\hline
			$Q$   & 3     & 1     & m \\
			$Q'$  & 1     & 3     & n \\
			\hline
		\end{tabular}%
		
		\label{tab: fermion classify}%
	\end{table}%

	The alignment axion model has been widely discussed in the literature~\cite{Kim2005,Choi2014a,Kappl2014a}. $SU(3)_h$ is introduced with a new free parameter of $CP$-violation $\theta'$. As shown in Fig.~\ref{fig:figure1}(a), massive fermions $Q(3,1)$ and $Q'(1,3)$ are charged under both $U(1)_{PQ}$ symmetry and $U(1)_{PQ'}$ symmetry. When $U(1)_{PQ}$ and $U(1)_{PQ'}$ broken, axion is left as a pseudo-Goldstone boson. The corresponding current and their divergence are 
	\begin{equation}\label{eq: Q current}
	\begin{aligned}
	J^\mu_{PQ}&=f_{a_1} \partial_{\mu}a_1+m_1\overline{Q}\gamma^\mu\gamma_5Q+n_1\overline{Q'}\gamma^\mu\gamma_5Q', \\
	J^\mu_{PQ'}&=f_{a_2} \partial_{\mu}a_2+m_2\overline{Q}\gamma^\mu\gamma_5Q+n_2\overline{Q'}\gamma^\mu\gamma_5Q', \\
	\partial_{\mu}J^\mu_{PQ}&=\frac{m_1g^{2}}{16 \pi^{2}} G_{a}^{\mu \nu} \tilde{G}_{a \mu \nu}+\frac{n_1{g'}^{2}}{16 \pi^{2}} {G'}_{A}^{\mu \nu} \tilde{G'}_{A \mu \nu}, \\
	\partial_{\mu}J^\mu_{PQ'}&=\frac{m_2g^{2}}{16 \pi^{2}} G_{a}^{\mu \nu} \tilde{G}_{a \mu \nu}+\frac{n_2{g'}^{2}}{16 \pi^{2}} {G'}_{A}^{\mu \nu} \tilde{G'}_{A \mu \nu}.
	\end{aligned}
	\end{equation} 
	where $G_{a}^{\mu \nu}$ and ${G'}_{A}^{\mu \nu}$ are strength tensors of gauge fields for QCD and hidden QCD respectively. $g$ and $g'$ are couplings of gauge interactions. In some papers, the hidden sector is extended to a ``mirrored SM'' with massive fermions $u'$ and $d'$ that do not carry $U(1)_{PQ}$ charge like u and d in the SM~\cite{Gherghetta2016a,Hook2014ASttSCP}.

	The main structure of the dynamical solution~\cite{Choi1985DA} is shown in Fig~\ref{fig:figure1}(b).  Similar to the aligned axion model, $SU(3)_h$ is introduced with a new $CP$ parameter $\theta'$. Two massless fermions $\psi$ and $\chi$ are introduced to absorb $CP$-violating phases $\theta$ and $\theta'$ through a $U(1)_A$ transformation. It's common to assume that $SU(3)_h$ will confine just like QCD at a scale $f_{\pi_h}$ which is much higher than QCD confining scale $f_\pi$. The related currents are
	\begin{equation}\label{eq: anomalous current1}
	\begin{aligned}
	J^\mu_{A}(\psi)&=\overline{\psi}\gamma^\mu\gamma_5\psi, \quad J^\mu_{A}(\chi)=\overline{\chi}\gamma^\mu\gamma_5\chi  \\
	\partial_{\mu}J^\mu_{A}(\psi)&=\frac{3g^{2}}{16 \pi^{2}} F_{a}^{\mu \nu} \tilde{F}_{a \mu \nu}+\frac{3g'^{2}}{16 \pi^{2}} G_{A}^{\mu \nu} \tilde{G}_{A \mu \nu}, \\
	\partial_{\mu}J^\mu_{A}(\chi)&=\frac{g'^{2}}{16 \pi^{2}} G_{A}^{\mu \nu} \tilde{G}_{A \mu \nu},
	\end{aligned}
	\end{equation}

Below the scale $f_{\pi_h}$, we assume the hidden sector has a dynamical chiral symmetry breaking caused by the fermion condensate with
	\begin{equation}\label{condensate}
	\langle\bar{\psi} \psi\rangle \approx - c_\psi f_{\pi_h}^{3}, \qquad \langle\bar{\chi} \chi\rangle \approx - c_\chi f_{\pi_h}^{3},
	\end{equation}
	where $c_\psi$,$c_\chi$ are constants and $f_{\pi_h} \gg f_\pi$.  Considering QCD as an additional ``flavor symmetry'', there is a $SU(4)_L\times SU(4)_R$ symmetry between $\psi$ and $\chi$. After the condensate of hidden QCD, $\pi_h^a$ are Goldstones corresponding to   coset generators of $SU(4)_L\times SU(4)_R/SU(4)_V$. The decomposition of $\pi_h^a$ into $SU(3)_c$ is
	\begin{equation}
	15=8+3+\overline{3}+1.
	\end{equation}
	The QCD color-singlet scalar is denoted as $\pi_h\equiv \pi_h^{15}$ for short. And the rest 14 colored scalars are supposed to be heavy because of QCD condensate. However, there is one more color-singlet scalar $\eta_{h}^{\prime}$ related to the $U(1)_A$ symmetry. The corresponding currents related to these two fields are
	\begin{equation}\label{current 12}
	\begin{aligned}
	J_\mu(\pi_h)&=\dfrac{1}{\sqrt{6}}(\overline{\psi}^c \gamma_\mu\gamma_{5} \psi_c-3\overline{\chi} \gamma_\mu\gamma_{5}\chi), \\
	J_\mu(\eta_{h}^{\prime})&=\dfrac{1}{\sqrt{2}}(\overline{\psi}^c \gamma_\mu\gamma_{5} \psi_c+\overline{\chi} \gamma_\mu\gamma_{5}\chi),
	\end{aligned}
	\end{equation}
	where `c' is the color index for $\psi$. These currents have been normalized and we have chosen $tr(T^aT^b)=2\delta^{ab}$ as normalization of $SU(4)$ generators. The specific matrices of $SU(4)$ generators and details of derivation are shown in Appendix~\ref{A1}. As a result of condensate, we can use $\pi_h$ and $\eta_{h}^{\prime}$ standing for the $U(1)_A$ symmetry at low energy scale. $\theta$ and $\theta'$ can be offseted when $\pi_h$ and $\eta_{h}^{\prime}$ take VEVs.


	
	

	\section{Cancellation of the Strong $CP$ Phase}
	\label{section3}
	
	When considering the extension of the SM, new contributions to the strong $CP$ violation are introduced. Therefore, the strong $CP$ phase $\overline{\theta}$ needs to be modified. For distinction, the modified strong $CP$ phase is notated as $\theta_{phy}$. In this section, we discuss these new contributions and methods to solve the strong $CP$ problem. The difficulty is that $\overline{\theta}$ is at order $ 1$ but $\theta_{phy}$ is at order $  10^{-10}$. Axion models and dynamic solutions both solve this difficulty by canceling $\theta_{phy}$ at the minimal point of goldstone potential, which links $\theta_{phy}$ with $\overline{\theta}$.  
	To calculate $\theta_{phy}$ which describes $CP$ violation effects at the scale lower than $\Lambda_{QCD}$, Chiral Perturbation Theory (ChPT) is used to match fields of quarks with fields of hadrons. Indeed, it is sufficient to consider ChPT with two light quarks $u$ and $d$. The Lagrangian for QCD with an axion at hadron level is~\cite{Kuster2008A}
	\begin{equation}
	\mathcal{L}_2=\dfrac{f_\pi^2}{4} \mathrm{Tr} \left( \partial_{\mu}U\partial^\mu U^\dagger\right) +A f_\pi^3 \mathrm{Tr} \left(MU^\dagger+UM^\dagger \right)+B f_\pi^4\left(\theta+ \frac{a}{f_a}+\frac{i}{2} \mathrm{Tr}\left(\log U-\log U^{\dagger}\right)\right)^2  ,
	\end{equation}
	with
	\begin{equation}
	U=\exp\left[  \dfrac{ i\left( \pi^a\tau^a+I \eta \right) }{f_\pi} \right], \qquad M=
	\begin{pmatrix}
	m_u & 0 \\
	0 & m_d
	\end{pmatrix}  ,
	\end{equation}
	where $\pi^a$, $\eta$ are mesons and $a$ is the aixon. $A$ and $B$ are dimensionless parameters matched by meson masses. Here $\tau^a$ are Pauli matrices and $I$ is the identity matrix. For convenience, we set the matrix of quark mass to be real in the rest of this paper, leading to $\theta=\overline{\theta}$. The potential part is shown as 
	\begin{equation}\label{eq: QCD potential}
	V = -2Af_\pi^3 \left[ m_u\cos(\frac{\pi^0+\eta}{f_\pi})+m_d\cos(\frac{\pi^0-\eta}{f_\pi})\right] +Bf_\pi ^4\left( \frac{2\eta}{f_\pi}+\theta+ \frac{a}{f_a}\right) ^2.
	\end{equation}
	For convenience, $\left\langle\pi^0\right\rangle/f_\pi$, $\left\langle\eta\right\rangle/f_\pi$ and $\left\langle a\right\rangle/f_a$ are defined as phases $\phi_1$, $\phi_2$ and $\phi_a$ respectively.
	Considering $CP$ transformation, $\pi^0$, $\eta$ and $a$ are changed into $-\pi^0$, $-\eta$ and $-a$. Once $CP$ is conserved in eq.~\eqref{eq: QCD potential}, these phases need to meet 
	\begin{equation}\label{$CP$ conserved1}
	\phi_1+\phi_2=0, \qquad \phi_1-\phi_2=0, \qquad 2\phi_2+\theta+\phi_a=0,
	\end{equation}
	which is equivalent to 
	\begin{equation}\label{$CP$ conserved2}
	\phi_1=\phi_2=0, \qquad \theta+\phi_a=0.
	\end{equation}
	Considering the derivative of the effective potential in eq.~\eqref{eq: QCD potential} with respect to $\pi^0$, $\eta$ and $a$, eq.~\eqref{$CP$ conserved2} is exactly the solution at the minimal point of the potential. New contributions from $U(1)$ symmetries are described by phase $\phi_a$. Finally, the $CP$-violating observable $\theta_{phy}$ is gotten as
	\begin{equation}\label{eq:theta}
	\theta_{phy}=\theta+\phi_a=\theta+\frac{\left\langle a \right\rangle}{f_a}.
	\end{equation}
	The strong $CP$ problem is solved by the offset of axion VEV and the phase $\theta$ to ensure $\theta_{phy}=0$. Axion mass can also be derived from eq.~\eqref{eq: QCD potential} as 
	\begin{equation}\label{axion mass}
	m_{a}^2 =m_{\pi}^2 \frac{f_\pi^2}{f_a^2}\frac{m_u m_d}{(m_u+m_d)^2}.
	\end{equation}
	From this QCD axion model, it can be found that $\theta_{phy}$ needs to be modified due to the new contribution from axion. Similar absorption happens in other axion models and dynamical solutions.

	For a general hidden QCD model with new fermions added, two UV free parameters, $ \theta$ and $ \theta'$, have to be introduced. In general cases, the potential term induced by the instanton effect is
	\begin{equation}\label{eq: theta_thetaprime potential}
	V_\theta = B_1f_\pi ^4\left( \sum_{i=1}^2 m_i \frac{a_i}{f_{a_i}}+\theta\right) ^2+B_2f_{\pi_h} ^4\left(  \sum_{i=1}^2 n_i \frac{a_i}{f_{a_i}}+\theta' \right) ^2,
	\end{equation}
	where $m$ and $n$ stand for different anomalous charges.
	$\footnote {It's more complex for dynamical solutions, because normalization factor also has to be included in this $m$ and $n$.}$
	
	Strong $CP$ problems for both QCD and hidden QCD can be solved only satisfying
	\begin{equation}\label{eq: theta_hid_QCD}
	\begin{aligned}
	\theta_{phy}&=\theta+  \frac{\left\langle \eta ' \right\rangle}{f_\pi}  + \theta_{U(1)} =0,\\
	\theta'_{phy}&=\theta'+\theta'_{U(1)} =0.
	\end{aligned}
	\end{equation}
	In the preceding equation, $ \left\langle \eta ' \right\rangle$ means the contribution from VEV of $\eta'$, which can be absorbed by redefinition of $\theta$. We are not interested in it, so we will absorb it with $\theta$ in the following discussion. $ \theta_{U(1)}$ and $ \theta'_{U(1)}$ come from axion VEVs. 
	In following subsections, we discuss the $\theta$ cancellation of eq.~\eqref{eq: theta_hid_QCD} in some specific models with different $U(1)$ symmetries.
	
	
	\subsection{One U(1)}
\begin{figure}[h]
		\centering
		\includegraphics[width=1\linewidth]{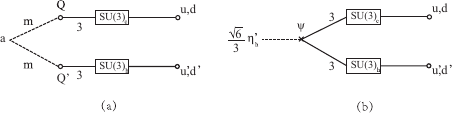}
		\caption{(a): Structure of $U(1)_{PQ}$ models with mirror symmetry. (b): Structure of $U(1)_{A}$ dynamic solutions with mirror symmetry.}
		\label{fig:oneu1}
	\end{figure}

	As shown in eq.~\eqref{eq: theta_hid_QCD}, there are two degrees of freedom needed to absorb both $\theta$ and $ \theta'$. When only one $U(1)$ is introduced, there is only one new degree of freedom. That means $\theta$ and $ \theta'$ cannot be absorbed simultaneously and the strong $CP$ problem cannot be solved. 
	To solve this problem, a mirrored $Z_2$ symmetry has to be introduced between $SU(3)_{c}$ in the SM and $SU(3)_h$ in the hidden sector.
	In UV theories, some models can achieve this by embedding these two $SU(3)$ into a larger gauge symmetry group $SU(6)$~\cite{Gherghetta2016a}. Consequently, two $CP$ phases are forced to be identical, which means
	\begin{equation}\label{Z_2}
	\theta= \theta' .
	\end{equation}
	As shown in Fig.~\ref{fig:oneu1}, the additional $U(1)$ could be achieved by two methods. One method introduces an axion with $U(1)_{PQ}$ (left) and the other uses massless $\psi(3,3)$ with $U(1)_{A}$ (right). There are detailed discussions about both two methods in literature~\cite{Rubakov1997,Berezhiani2001a,Hook2014ASttSCP,Fukuda2015AMoVQA,Albaid2015SCaS,Chiang2016}. For the left one, cancellation eq.~\eqref{eq: theta_hid_QCD} becomes
	\begin{equation}
	\theta_{phy}^{\prime}= \theta_{phy}=\theta+\frac{\left\langle a\right\rangle}{f_a}=0 .
	\end{equation}
	For the right one, as shown in Appendix~\ref{A1}, the cancellation eq.~\eqref{eq: theta_hid_QCD} is deduced to
	\begin{equation}
	\theta_{phy}^{\prime}= \theta_{phy}= \theta+\sqrt{6}\dfrac{\left\langle \eta'_{h} \right\rangle }{f_{\pi_h}} =0 .
	\end{equation}
	In both cases, the strong $CP$ problem will be solved by only one $U(1)$ symmetry. And the $CP$-violating effects from $\theta$ and $\theta'$ will be counteracted by $ \left\langle a\right\rangle$ or $ \left\langle \eta'_{h} \right\rangle $.
	
	\subsection{Two Similar U(1)s}
	
	As in the previous subsection, models with two $U(1)$ freedoms can also be sorted by the different mechanism of $U(1)$ symmetries. As shown in Fig.~\ref{fig:figure1}, models could contain two $U(1)_{PQ}$ or two $U(1)_{A}$ symmetries. For these kinds of models, axial currents are determined by representations of fermions in $SU(3)_c$ and $SU(3)_h$. The cancellation equation of $\theta$ and $\theta'$ will be determined by these currents and different $U(1)$ charge for fermions.

	Here shows another advantage of our diagram method that the cancellation equation can be read easily. Fig.~\ref{fig:figure1}(a) shows models with two $U(1)_{PQ}$ symmetries. Choosing $SU(3)_c$ as the starting point responsible for $\theta$, we can find all possible links end at left side. Notice these links can only go from right to left. The cancelling phase of each link is the product of the charge number on the dashed line and the scale factor at the end point on the left. Finally we can sum all offset terms from different links together. For $\theta'$, the starting point will be changed into $SU(3)_h$, then repeat these steps. The cancellation equations from our diagram read
	\begin{equation}\label{eq: theta_axion_}
	\begin{aligned}
	\theta_{phy}&=\theta+ m_1\dfrac{\left\langle a_1\right\rangle }{f_{a_1}} +m_2\dfrac{\left\langle a_2\right\rangle }{f_{a_2}}=0, \\
	\theta_{phy}^{\prime}&=\theta'+ n_1\dfrac{\left\langle a_1\right\rangle }{f_{a_1}}+n_2\dfrac{\left\langle a_2\right\rangle }{f_{a_2}}=0 .
	\end{aligned}
	\end{equation}
	This two-axion model is widely discussed. When the angle between vectors $(m_1,m_2)$ and $(n_1,n_2)$ is near zero (but not zero) on the contrast to the large m and n, it is exactly the `alignment axion' model~\cite{Kim2005,Choi2014a}.
	
	Models with two $U(1)_A$ symmetries provided by massless fermions have similar expressions for $CP$-violating angles. In Fig.~\ref{fig:figure1}(b), $M_\psi=0$ and $M\chi=0$ are supposed. The corresponding currents has been given in eq.~\eqref{current 12}. It's a little more complex to read cancellation equations for this model from the figure. There will be two more things to notice. First, the normalization factor has been already written together with scale factor. Second, the degeneracy should be multiplied for each link, which comes from the color index of the other $SU(3)$ outside the link. The degeneracy can be read from the number for representation on other solid lines attached to the fermion field in the link. For example, the degeneracy of the link, ``$SU(3)_c \rightarrow \psi \rightarrow \pi_h$", will be the number `3' on the solid line attached $\psi$ and $SU(3)_h$. When sum all links together, this link should be multiplied by 3. The final cancellation equations are
	\begin{equation}
	\begin{aligned}
	\theta_{phy}&=\theta+ \frac{\sqrt{6}}{2}\dfrac{\left\langle \pi_h\right\rangle }{f_{h}} +\frac{3\sqrt{2}}{2}\dfrac{\left\langle \eta'_{h}\right\rangle }{f_{h}}=0, \\
	\theta_{phy}^{\prime}&=\theta'+ 2\sqrt{2}\dfrac{\left\langle \eta'_{h}\right\rangle }{f_{h}}=0 .
	\end{aligned}
	\end{equation}
	The cancellation equations of other figures can also be read in this way.
	
	\subsection{Two Different U(1)s}
	
\begin{figure}[h]
		\centering
		\includegraphics[width=1\linewidth]{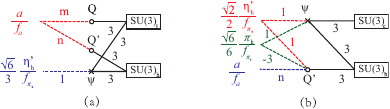}
		\caption{Structure of models with mixed $U(1)$s. (a) Model A: the spontaneous breaking of $U(1)_{PQ}$ is independent with hidden QCD. (b) Model B: the dynamical symmetry breaking of hidden QCD induces $U(1)_{PQ}$ breaking.}
		\label{fig: two U_1}
	\end{figure}

	In these two-$U(1)$ solutions, $U(1)$s need not to be the same, which means models of one $U(1)_{PQ}$ and one $U(1)_A$ is possible. Here we show two simple examples of that new kind of models in Fig.~\ref{fig: two U_1}. 
	
	The mixed solution of $U(1)_{PQ}$ and $U(1)_A$ would be more complex, for the reason that two independent scales, $f_{h}$ and $f_a$, are simultaneously involved. $f_{h}$ is the condensate scale of hidden QCD fermions, while $f_a$ is the spontaneously breaking scale of the Peccei-Quinn symmetry. Fig.~\ref{fig: two U_1}(a) shows model A in which $U(1)_{PQ}$ is spontaneous broken independently with the hidden QCD. The result of cancellation equations in this model is
	\begin{equation}
	\begin{aligned}
	\theta_{phy}&=\theta+\sqrt{6}\dfrac{\left\langle \eta'_{h} \right\rangle }{f_{h}}+ m\dfrac{\left\langle a\right\rangle }{f_a}=0, \\
	\theta_{phy}^{\prime}&=\theta'+\sqrt{6}\dfrac{\left\langle \eta'_{h} \right\rangle }{f_{h}}+ n\dfrac{\left\langle a\right\rangle }{f_a}=0 .
	\end{aligned}
	\end{equation}
	Fig.~\ref{fig: two U_1}(b) shows model B in which the spontaneous breaking of $U(1)_{PQ}$ is induced by the hidden QCD . The cancellation equations are
	\begin{equation}
	\begin{aligned}
	\theta_{phy}&=\theta+\dfrac{3}{\sqrt{2}}\dfrac{\left\langle \eta'_{h} \right\rangle }{f_{h}}+\dfrac{3}{\sqrt{6}}\dfrac{\left\langle \pi_h \right\rangle }{f_{h}}=0, \\
	\theta_{phy}^{\prime}&=\theta'+2\sqrt{2}\dfrac{\left\langle \eta'_{h} \right\rangle }{f_{h}}+ n\dfrac{\left\langle a\right\rangle }{f_a}=0 .
	\end{aligned}
	\end{equation}
	It seems that redundancy exists in these equations, as there are three VEVs to absorb two $\theta$ parameters. However, there is an additional equation constraining these three VEVs from the Yukawa term of $Q'$, as shown in~\eqref{L_mass} in the next section.

	
	\section{Physical Observables}
	\label{section4}
	In this section, we study the most important physical observables, axion mass and axion-photon coupling in the one-axion solution and mixed two-axion solutions, which can not directly be acquired from diagrams in the last section. We compare the axion masses in these two kinds of solutions. We also discuss the relation between the physical axion decay constant $F_a$ and  the axion-photon coupling $g_{a\gamma\gamma}$ in the mixed two-axion solution.
	 
	\subsection{One-axion Solution}
	The one axion solution is exactly the implementation of  the visible axion model in which axion mass is enhanced by the condensate scale of the hidden QCD. We illustrate this with a specific model shown in  Fig.~\ref{fig:oneu1}(a). In this model, there is only one $U(1)_{PQ}$ chiral symmetry. The corresponding current and its divergence are
	\begin{equation}
	J^\mu_{PQ}=f_a \partial_{\mu}a+m\overline{Q}\gamma^\mu\gamma_5Q+m\overline{Q'}\gamma^\mu\gamma_5Q', \qquad
	\partial_{\mu}J^\mu_{PQ}=\frac{mg^{2}}{16 \pi^{2}} F_{a}^{\mu \nu} \tilde{F}_{a \mu \nu}+\frac{mg'^{2}}{16 \pi^{2}} G_{A}^{\mu \nu} \tilde{G}_{A \mu \nu}.
	\end{equation}
	At the low energy scale, the axion mass term stems from the instanton effects of QCD and the hidden QCD. Therefore, the mass terms are
	\begin{equation}
	-\mathcal{L}_{\text{mass}}=\cdots + \Lambda_c^4 \left(\frac{ma}{f_{a}}\right)^2+\Lambda_h^4 \left(\frac{ma}{f_{a}}\right)^{2},
	\end{equation}
	where $\Lambda_c$ and $\Lambda_h$ denote the confinement scale of QCD and the hidden QCD respectively. Due to the contribution from the high energy scale $\Lambda_h$, the axion mass could be much heavier than KSVZ or DFSZ axion. Different phenomena of the heavy axion may appear in collider experiments or rare decay of mesons~\cite{Gherghetta2016a}.
	\subsection{Two-axion Solution}
	In two-axion solutions with extra hidden QCD, the light mass eigenstate always exists and can be treated as an invisible axion, even if $\Lambda_h$ is much higher than QCD scale $\Lambda_c$. We illustrate this with the mass matrix of axions by a perturbative method. The Lagrangian of the mass part can be expressed as
	\begin{equation}\label{L_two_axions_mass}
	-\mathcal{L}_{\text{mass}}=\cdots + \Lambda_c^4 \left(\frac{m_1a_1}{f_{a_1}}+\frac{m_2a_2}{f_{a_2}}\right)^2+\Lambda_h^4 \left(\frac{n_1a_1}{f_{a_1}}+\frac{n_2a_2}{f_{a_2}}\right)^{2},
	\end{equation}
	where $f_{a_1}$ and $ f_{a_2}$ are scale factors. Anomaly  coefficients of axions are denoted as $m_1,\ m_2,\ n_1$ and $n_2$. The corresponding mass matrix can be expressed as
	\begin{equation}
	\Lambda_c^4\begin{pmatrix}
	m_1^2\frac{1}{f_{a_1}^2}	&m_1m_2\frac{1}{f_{a_1}f_{a_2}} \\ 
	m_1m_2\frac{1}{f_{a_1}f_{a_2}}	& m_1^2\frac{1}{f_{a_2}^2}
	\end{pmatrix}
	+
	\Lambda_h^4\begin{pmatrix}
	n_1^2\frac{1}{f_{a_1}^2}	&n_1n_2\frac{1}{f_{a_1}f_{a_2}} \\ 
	n_1n_2\frac{1}{f_{a_1}f_{a_2}}	& n_1^2\frac{1}{f_{a_2}^2}
	\end{pmatrix} .
	\end{equation}
	Since $\Lambda_h \gg \Lambda_c$, the first matrix with respect to $\Lambda_c $ can be treated as perturbation. Considering zero order approximation, the heavy degree of freedom is denoted as
	\begin{equation}\label{A}
	A \propto \frac{n_1a_1}{f_{a_1}}+\frac{n_2a_2}{f_{a_2}},
	\end{equation}
	while the massless one is in the orthogonal direction with $A$ in field space, which is
	\begin{equation}\label{a}
	a \propto \frac{n_2a_1}{f_{a_2}}-\frac{n_1a_2}{f_{a_1}}.
	\end{equation}
	Replacing $(a_1,\ a_2)$ with $(A,\ a)$, eq.~\eqref{L_two_axions_mass} becomes
	\begin{equation}\label{diagonalized}
	-\mathcal{L}_{\text{mass}}=\cdots + \Lambda_c^4 \left(\cos\alpha\frac{a}{F_a}+\sin\alpha\frac{A}{F_A}\right)^2+\Lambda_h^4 \left(\frac{A}{F_A}\right)^{2},
	\end{equation}
	where $F_A$ ($F_a$) is the decay constant of heavy axion ( light axion) after diagonalization and normalization. $\alpha$ is the mixing angle. In general, $F_a$ and $F_A$ are comparable. Therefore, the heavier one has a mass of $m_A^2 \sim \Lambda_h^4/F_A^2$ and the lighter one has a mass of $m_a \sim \Lambda_c^4/F_a^2$.
	The lighter axion is exactly the invisible QCD axion.
	
	In the following part, we discuss two specific models of different breaking mechanisms in mixed two-axion solutions. The model with two independently breaking $U(1)$ symmetries (model A) is shown in Fig.~\ref{fig: two U_1}(a)  and  the model with one breaking $U(1)$ symmetry induced by the other (model B) is shown in Fig.~\ref{fig: two U_1}(b). The main difference between these two models is whether the symmetry breaking of $U(1)_{PQ}$ is induced by the breaking of $U(1)_A$.
	Chiral symmetries are same in both cases. The corresponding currents and its divergence of these models are
	\begin{equation}\label{eq: anomalous current}
	\begin{aligned}
	J^\mu_{A}&=\dfrac{\sqrt{6}}{3}\overline{\psi}\gamma^\mu\gamma_5\psi, \qquad J^\mu_{PQ}=f_a \partial_{\mu}a+(m\overline{Q}\gamma^\mu\gamma_5Q)+n\overline{Q'}\gamma^\mu\gamma_5Q', \\
	\partial_{\mu}J^\mu_{A}&=\frac{\sqrt{6}g^{2}}{16 \pi^{2}} G_{a}^{\mu \nu} \tilde{G}_{a \mu \nu}+\frac{\sqrt{6}g'^{2}}{16 \pi^{2}} {G'}_{A}^{\mu \nu} \tilde{G'}_{A \mu \nu}, \\
	\partial_{\mu}J^\mu_{PQ}&=\frac{mg^{2}}{16 \pi^{2}} G_{a}^{\mu \nu} \tilde{G}_{a \mu \nu}+\frac{ng'^{2}}{16 \pi^{2}} {G'}_{A}^{\mu \nu} \tilde{G'}_{A \mu \nu},
	\end{aligned}
	\end{equation}
	where we set the anomaly coefficient $m=0$ in the second model for simplicity.
	\subsubsection{Model A}
	\label{indenpent}
	In model A, the spontaneous breaking of two chiral $U(1)$ symmetries comes from different mechanisms. The $U(1)_{PQ}$ symmetry is broken by the effective potential of the complex scalar $\phi$, while $U(1)_{A}$ is broken by the condensate of massless fermion $\psi$. Therefore, these two scales are independent. The only source of explicit breaking comes from the instanton effects contributing to masses of both axions. The mass-relevant Lagrangian is
	\begin{equation}\label{eq:L_mass}
	-\mathcal{L}_{\text{mass}}=\cdots + bf_\pi^4 \left(\frac{\sqrt{6}\eta^{\prime}_h}{f_{\pi_h}}+\frac{ma}{f_{a}}\right)^2+b'f_{\pi_h}^4 \left(\frac{\sqrt{6}\eta^{\prime}_h}{f_{\pi_h}}+\frac{na}{f_{a}}\right)^{2},
	\end{equation}
	where the first(second) term stems from QCD(hidden QCD) instanton effects with $f_{\pi_h} \gg f_\pi$.  After diagonalizing the mass matrix, we find the lighter mass eigenstate $a_{phy}$ to be
	\begin{equation}\label{axion mass 1}
	a_{\text{phy}}=\dfrac{nf_{\pi_h} \eta^{\prime}_h-\sqrt{6} f_{a} a}{\sqrt{{6f^2_a+n^2}f_{\pi_h}^2}},
	\end{equation}
    with its mass and decay constant are
	\begin{equation}\label{F_a}
	m_{a_\text{phy}}^2F_{a_\text{phy}}^2 =m_{\pi}^2 f_\pi^2\frac{m_u m_d}{(m_u+m_d)^2}, \qquad F_{a_\text{phy}}=\dfrac{\sqrt{{6f^2_a+n^2}f_{\pi_h}^2}}{\sqrt{6}|m-n|}.
	\end{equation}
	Furthermore, the axion-photon coupling $g_{{a_{phy}} \gamma \gamma}$ can be defined as
	\begin{equation}\label{eq:L_arr}
	\begin{aligned}
	\mathcal{L}_{{a_{phy}} \gamma \gamma}=\frac{1}{4}g_{{a_\text{phy}} \gamma \gamma}a F_{\mu \nu} \tilde{F}^{\mu \nu}=\frac{1}{4} (g_{{a_\text{phy}} \gamma \gamma}^\text{IR}+g_{{a_\text{phy}} \gamma \gamma}^\text{UV})a F_{\mu \nu} \tilde{F}^{\mu \nu} ,
	\end{aligned}
	\end{equation}
	where $g_{{a_\text{phy}}\gamma \gamma}^\text{IR}$ comes from the mixing between axion and $\pi^0$.  The latest result is accurate to NLO~\cite{Cortona2015}, which is 
	\begin{equation}
 g_{{a_\text{phy}} \gamma \gamma}^\text{IR}=-1.92(4) \frac{\alpha_{\mathrm{em}}}{2 \pi F_{a_\text{phy}}}.
	\end{equation}
	The contribution from UV part is model-dependent. In the QCD axion model, this contribution is
	\begin{equation}\label{qcdcase}
	 g_{{a_\text{phy}} \gamma \gamma}^\text{UV}=\dfrac{E}{N} \frac{\alpha_{\mathrm{em}}}{2 \pi F_{a_\text{phy}}},
	\end{equation}
where E and N are respectively  Electromagnetic and the color anomaly coefficients. In the model A, $g_{a \gamma \gamma}^\text{UV}$ has contribution from the mixing of $\eta'_h$ and $a$. the interactions by anomaly effect between  $\eta'_h$, $a$ and photon can be written as
	\begin{equation}\label{garr1}
	\mathcal{L}_{a\gamma \gamma}=\left(\sqrt 6q^2_\psi \frac{\eta^{\prime}_h}{f_{\pi_h}}+6(q_Q^2+q_{Q'}^2)\frac{a}{f_{a}} \right) \frac{\alpha_{\mathrm{em}}}{4\pi}F_{\mu \nu} \tilde{F}^{\mu \nu},
	\end{equation}
	where $q_\psi$, $q_Q$ and $q_{Q'}$ denote the electric charge of each particles. Therefore, 
	\begin{equation}\label{coupling}
g_{{a_\text{phy}}\gamma\gamma}^\text{UV}=\dfrac{nq^2_\psi-6(q_Q^2+q_{Q'}^2)}{2|m-n|}\frac{\alpha_{\mathrm{em}}}{2 \pi F_{a_\text{phy}}}.
	\end{equation}
	When setting $n\gg|m-n|$, the UV contribution of the axion-photon coupling is enlarged. Meanwhile the physical axion decay constant meets the condition $F_{a_\text{phy}} \gg f_a $ as shown in eq.~\eqref{F_a}. This result is actually a special case of `alignment axion' models~\cite{Kim2005,Choi2014a}.
	
	\subsubsection{Model B}
	\label{induced}
	In model B, the spontaneous breaking of $U(1)_{PQ}$ is triggered by the condensate of the light fermion $Q'$ in the hidden sector. Thus two breaking scales are related. We assume the potential of the complex scalar $\phi$ containing only a quadratic term with $\mu^2 > 0$. The Yukawa term of $Q'$ will induce a tadpole term of $\phi$ after the chiral symmetry breaking in the hidden sector. The Lagrangian after the chiral symmery breaking is
	\begin{equation}
	\begin{aligned}\label{Lag}
	\mathcal{L}_\text{CSB}&=\partial_{\mu}\phi\partial^{\mu}\phi^*-\mu^2\phi^2
	+\dfrac{f_{\pi_h}^{2}}{4} \text{Tr} \partial_{\mu} \Sigma \partial^{\mu} \Sigma^{\dagger}+a' f_{\pi_h}^{3} \text{Tr}\left(H\Sigma^{\dagger}+H^{\dagger} \Sigma\right)+\cdots, \\
	\end{aligned}
	\end{equation}
	where
	\begin{equation}
	\Sigma=\exp\left[  \dfrac{ i\left( \pi_h^aT^a+\frac{1}{\sqrt{2}}I_{4\times 4} \eta'_h \right) }{f_\pi} \right] ,
	\qquad
	H=
	\begin{pmatrix}
	0 &  &  &  \\ 
	& 0 &  &  \\ 
	&  & 0 &  \\ 
	&  &  & y\phi
	\end{pmatrix} .
	\end{equation}
	Here $T^a$ are generators of the $SU(4)$ group meeting the trace condition $\text{Tr}(T^aT^b)=2\delta^{ab} $. The anomalous term is omitted in eq.~\eqref{Lag}. Matrix $H$, the Yukawa terms of $\psi$ and $Q'$, is treated as a spurion field transforming as an adjoint representation under the $SU(4)_V$. The third term in eq.~\eqref{Lag} is the linear term of $\phi$ which triggers $U(1)_{PQ}$ breaking. Specifically, the complex scalar $\phi$ is parameterized as $\phi=\rho e^{i\left( na/{f_a} \right)} $. The effective potential of $\rho$ and $a$ is 
	\begin{equation}\label{potential phi}
	V(\rho,a)=\mu^{2} \rho^{2}-2a' y\rho f_{\pi_h}^{3} \cos \left(n\frac{a}{f_a}+\cdots \right),
	\end{equation}
	where $``\cdots"$ omits contributions from other composite scalars. Then we can estimate the VEV of $\rho$ to be $\left\langle \rho \right\rangle=f_a/\sqrt{2}  \sim 2y\dfrac{a'f_{\pi_h}^3}{\mu^2} $. Utilizing the  anomalous currents in eq.~\eqref{eq: anomalous current} and currents of $\pi_h$ and $\eta'_{h}$ in eq.~\eqref{current 12}, the mass terms of scalars in Lagrangian eq.~\eqref{Lag} becomes
	\begin{equation}\label{L_mass}
	-\mathcal{L}_{\text{mass}}=bf_\pi^4\left[ \sqrt{6}\left(\dfrac{1}{2}\dfrac{\pi_h}{f_{\pi_h}}
	+\dfrac{\sqrt{3}}{2}\dfrac{\eta^{\prime}_h}{f_{\pi_h}}\right)\right]^2+
	a'y\dfrac{f_af_{\pi_h}^{3}}{\sqrt{2}}\left(\dfrac{na}{f_a}-\dfrac{3}{\sqrt{6}}\frac{\pi_h}{f_{\pi_h}}+\dfrac{1}{\sqrt{2}} \frac{\eta_{h}^{\prime}}{f_{\pi_h}} \right)^2
	+
	b'f_{\pi_h}^4 \left(\frac{2\sqrt{2}\eta^{\prime}_h}{f_{\pi_h}}\right)^{2},
	\end{equation}
	where the second term stems from the Yukawa term of $Q'$.
	The $na/f_a$ in second term could move to the third term by a chiral rotation $\theta' \rightarrow \theta'+na/f_a $ and generates additional differential interaction of $a$.
	By means of the same steps in the former subsection, the lightest mass eigenstate is 
	\begin{equation}\label{axion mass 2}
	a_{phy}=\dfrac{nf_{\pi_h} \pi_h+3/\sqrt{6} f_{a} a}{\sqrt{{3/2f^2_a+n^2}f_{\pi_h}^2}}, \qquad F_{a_\text{phy}}=\dfrac{2\sqrt{3/2f^2_a+n^2f_{\pi_h}^2}}{\sqrt{6}|n|}.
	\end{equation}
	The $g_{a\gamma\gamma}^{\text{UV}}$ coming from $\pi_h$ to $\gamma\gamma$ is
	\begin{equation}\label{garr2}
	\mathcal{L}_{\pi_h\gamma\gamma}=\dfrac{\sqrt{6}}{2}(q^2_\psi-q^2_{Q^{\prime}}) \frac{\pi_h}{f_{\pi_h}}  \frac{\alpha_{\mathrm{em}}}{4\pi}F^{\mu \nu} \tilde{F}_{\mu \nu}.
	\end{equation}
	Therefore, the UV part contribution of light axion is
	\begin{equation}\label{coupling2}
g_{{a_\text{phy}}\gamma\gamma}^\text{UV}=2(q^2_\psi-q^2_{Q^{\prime}})\frac{\alpha_{\mathrm{em}}}{2 \pi F_{a_\text{phy}}}.
	\end{equation}

	In this model, the axion-photon coupling $g_{{a_\text{phy}} \gamma \gamma}$ is independent of the anomaly coefficient $n$, which is different from the QCD axion~\eqref{qcdcase} and normal two-axion models~\eqref{coupling}. The decay constant $F_{a_\text{phy}}$ is inversely proportional to $|n|$ . Furthermore, there is another colorless scalar $\pi_h$ with $m^2_{\pi_h} \sim f_af_{\pi_h}$.
	
	\section{ Conclusion and Discussion}
	\label{section5}
	We investigate all possible solutions to the strong $CP$ problem with anomalous $U(1)$ global symmetries when the hidden QCD is introduced, which include both the Peccei-Quinn mechanism and dynamical solutions. We find these two mechanisms can exist together, and a new class of solutions for the strong CP problem are proposed. Thus we classify two $U(1)$ models in an unified way, and discuss both one-$U(1)$ and two-$U(1)$ solutions in this work. 
	
	A ``moose-like'' diagram method is used to illustrate how the strong CP problem is solved via cancellation between CP phases, which could be read  directly. When the gauge interaction and new particles are given, a diagram can be determined with a breaking mode of $U(1)$ symmetries. From these diagrams, cancellation between $CP$ phases can be easily read without detailed analysis of the model. The information is important for solving the strong $CP$ problem. This method could be used to find new types of solutions and easily extended to cases of multi-$U(1)$ and multi-gauge interaction.
	
	In one-axion solutions, additional $Z_2$ symmetry keeps $\theta=\theta'$ to solve the strong $CP$ problem. 
	In two-axion solutions, two chiral $U(1)$s (either $U(1)_{PQ}$ or $U(1)_A$) have to be introduced. As a consequence, there always exists an invisible light axion and a heavy axion.

	Using the diagram method, we find two new mixed solutions. In Model A, the spontaneous breaking of $U(1)_{PQ}$ symmetry is triggered by the scalar potential, independent of the confinement of the hidden QCD, and the axion-photon couplings $g_{{a_\text{phy}}\gamma\gamma}^{UV}$ and the light axion decay constant $F_{a_\text{phy}}$ are both inversely proportional to the difference of anomalous charges $|m-n|$. In Model B, the  $U(1)_{PQ}$ is induced by the confinement of the hidden QCD, and  $g_{{a_\text{phy}}\gamma\gamma}^{UV}$ is independent of the anomaly coefficient $n$ of $Q'$, which is different with the QCD axion and normal two-axion models.
	The decay constant $F_{a_\text{phy}}$ is inversely proportional to $|n|$.	At low energy scales, it is hard to distinguish this light axion from the QCD axion, even extending to multi-$U(1)$ symmetries and multi-gauge interactions. However, this kind of solutions could be probed by detecting particles at the hidden QCD scale. Those heavy particles may leave some hints in cosmology and at the colliders. 
   
	\acknowledgments
We would like to thank Qing-Hong Cao and Shou-hua Zhu for valuable discussions and their support. 
D.H. is supported in part by the National Science Foundation of China under Grants No. 11635001, 11875072.
H.R.J is supported in part by the National Science Foundation of China under Grant Nos. 11725520, 11675002, 11635001.
H.L.L and J.H.Y. are supported by the National Science Foundation of China (NSFC) under Grants No. 11875003.  
M.L.X. is supported by the National Natural Science Foundation of China (NSFC) under grant No.2019M650856 and the 2019 International Postdoctoral Exchange Fellowship Program.
J.H.Y. is also supported by the National Science Foundation of China (NSFC) under Grants No. 11947302.

	\appendix
	\section{Dynamical Symmetry Breaking in Hidden Sector}\label{A1}
	 This appendix shows the details of dynamical symmetry breaking in hidden sector. Two specific examples in Fig.~\ref{fig:figure1}(b) and Fig.~\ref{fig:oneu1}(b) are discussed. The $su(N)$-algebra is also shown in this appendix.
	
	In dynamical solutions, the global symmetries in the hidden sector are broken as
	\begin{equation}
	SU(N)_L\times SU(N)_R \times U(1)_A \times U(1)_V \rightarrow SU(N)_V \times U(1)_V ,
	\end{equation}
 	where $U(1)_A$ is explicitly broken by instanton effects. The effective theory at low energy is assumed to be
	\begin{equation}
	\mathcal{L}=\dfrac{f_\pi^2}{4} \mathrm{Tr} \left( \partial_{\mu}U\partial^\mu U^\dagger\right)+B' f_\pi^4\left(\theta'+\frac{i}{2} \mathrm{Tr}\left(\log U-\log U^{\dagger}\right)\right)^2  ,
	\end{equation}
		with
	\begin{equation}
	U=\exp\left[  \dfrac{ i\left( \pi_h^aT^a+I \eta'_h \right) }{f_{\pi_h}} \right],
	\end{equation}
where $\pi_h^a$ are Goldstones corresponding to the $SU(N)_A$ symmetry and $\eta'_h$ is a pseudo Goldstone corresponding to the $U(1)_A$ symmetry. $T^a$ are basis elements of the $su(N)$-algebra and $I$ is the $N$-dimension identity matrix. We choose $tr(T^aT^b)=2\delta^{ab}$ as the normalization condition. Therefore $T^a$ are shown as
\begin{equation}
\begin{aligned}
T_1&=\left(\begin{array}{ccccc}
0 & 1 & 0  &  &0\\ 
1 & 0 & 0 & \vdots  & 0  \\ 
0 & 0 & 0& & 0 \\ 
 & \cdots & & \ddots  &\vdots \\
0 &0 &0 &\cdots  &0 
\end{array}\right), \quad
T_2=\left(\begin{array}{ccccc}
0 & -i & 0  &  &0\\ 
i & 0 & 0 & \vdots  & 0  \\ 
0 & 0 & 0& & 0 \\ 
& \cdots & & \ddots  &\vdots \\
0 &0 &0 &\cdots  &0 
\end{array}\right),\quad
T_3=\left(\begin{array}{ccccc}
1 & 0 & 0  &  &0\\ 
0 & -1 & 0 & \vdots  & 0  \\ 
0 & 0 & 0& & 0 \\ 
& \cdots & & \ddots  &\vdots \\
0 &0 &0 &\cdots  &0 
\end{array}\right),
\\
T_4&=\left(\begin{array}{ccccc}
0 & 0 & 1  &  &0\\ 
0 & 0 & 0 & \vdots  & 0  \\ 
1 & 0 & 0& & 0 \\ 
& \cdots & & \ddots  &\vdots \\
0 &0 &0 &\cdots  &0 
\end{array}\right), 
\quad
\cdots,
\quad   \!
T_{N^2-1}=\sqrt{\dfrac{2}{N(N-1)}}\left(\begin{array}{ccccc}
1 & 0 & 0  &  &0\\ 
0 & 1 & 0 & \vdots  & 0  \\ 
0 & 0 & 1& & 0 \\ 
& \cdots & & \ddots  &\vdots \\
0 &0 &0 &\cdots  &1-N
\end{array}\right).
\end{aligned}
\end{equation}
In general, scalars corresponding to diagonal generators and identity matrix ought to be mixed. $\eta'_{h}$ could be treated as axion when instanton effects are the only source that explicitly breaks $U(1)_A$. Therefore, scalars corresponding to diagonal generators and identity matrix are important to solve the strong CP problem. Specific forms of diagonal generators and identity matrix are
\begin{equation}\label{diagonal}
\begin{aligned}
T_3&=\diag(1, -1, 0 , 0,\cdots)\\
T_8&=\dfrac{1}{\sqrt{3}}\diag(1,1,-2,0,\cdots) \\
T_{15}&=\dfrac{1}{\sqrt{6}}\diag(1,1,1,-3, \cdots)\\
T_{N^2-1}&=\sqrt{\dfrac{2}{N(N-1)}}\diag(1,1,1,\cdots, 1-N)\\
I&=\dfrac{1}{\sqrt{N}}\diag(1,1,1,\cdots,1)
\end{aligned}
\end{equation}

In the model shown in Fig.~\ref{fig:figure1}(b), vectorlike fermions $\psi(3,3)$ and $\chi(1,3)$ are introduced. In this case, $\Psi=(\psi_1^C,\psi_2^C,\psi_3^C,\chi^C)$ has a $SU(4)_L\times SU(4)_R$ flavor symmetry in the hidden sector, under which $\Psi$ transforms as:
\begin{eqnarray}
\Psi\to e^{i(\theta^L_a T^a P_L + \theta^R_a T^a P_R)}\Psi = e^{i(\theta^V_a T^a+ \theta^A_a T^a \gamma^5)}\Psi.
\end{eqnarray}
For $SU(4)$, diagonal generators are $T_3,\ T_8$ and $T_{15}$.
After the condensate of hidden QCD, $\pi_h^a$ are Goldstones corresponding to   coset generators of $SU(4)_L\times SU(4)_R/SU(4)_V$. The decomposition of $\pi_h^a$ into $SU(3)_c$ is
\begin{equation}
15=8+3+\bar{3}+1.
\end{equation}
Colorless scalars are $\pi^{15}_h$ and $\eta'_{h}$ corresponding to $\gamma^5\otimes T_{15}$ and $\gamma^5\otimes I$ respectively. The rest 14 colored scalars are supposed to be heavy because of QCD condensate. Currents corresponding to these generators and their divergences are
\begin{equation}
\begin{aligned}
J_\mu(\pi^{15}_h)&=\overline{\Psi}\gamma_\mu\gamma_{5}T^{15}\Psi, \\
J_\mu(\eta'_{h})&=\overline{\Psi}\gamma_\mu\gamma_{5}I_{4\times4}\Psi, \\
\partial_{\mu}J^\mu(\pi^{15}_h)&=\dfrac{3}{\sqrt{6}}\frac{g'^{2}}{16 \pi^{2}} G_{a}^{\mu \nu} \tilde{G}_{a \mu \nu}, \\
\partial_{\mu}J^\mu(\eta'_{h})&=
2\sqrt{2}\frac{g'^{2}}{16 \pi^{2}} G_{a}^{\mu \nu} \tilde{G}_{a \mu \nu}+\dfrac{3}{\sqrt{2}}\frac{g^{2}}{16 \pi^{2}} {G'}_{A}^{\mu \nu} \tilde{G'}_{A \mu \nu}.
\end{aligned}
\end{equation}
Matching with these currents, we can get the potential of these composite axions at low energy as
\begin{equation}
V= B_1f_\pi ^4\left(\dfrac{3}{\sqrt{6}} \dfrac{\pi_{h}}{f_{\pi_h}} + 2\sqrt{2}\dfrac{\eta'_{h}}{f_{\pi_h}} +\theta' \right) ^2+ B_2f_{\pi_h} ^4\left(\dfrac{3}{\sqrt{2}} \dfrac{\eta'_{h}}{f_{\pi_h}} +\theta  \right) ^2.
\end{equation}
At the minimal point of the potential, cancellation equations of $CP$ phases are
\begin{equation}\begin{array}{l}
\theta_{p h y}=\theta+\dfrac{\sqrt{6}}{2} \dfrac{\left\langle\pi_{h}\right\rangle}{f_{h}}+\dfrac{3 \sqrt{2}}{2} \dfrac{\left\langle\eta_{h}^{\prime}\right\rangle}{f_{h}}=0 \\
\theta_{p h y}^{\prime}=\theta^{\prime}+2 \sqrt{2} \dfrac{\left\langle\eta_{h}^{\prime}\right\rangle}{f_{h}}=0
\end{array}\end{equation}

In the model shown in Fig.~\ref{fig:oneu1}(b),The fermion $\psi(3,3)$ is introduced. In this case, $N=3$, $\Psi=(\psi_1^C,\psi_2^C,\psi_3^C)$. The only colorless scalar is $\eta'_{h}$ corresponding to $I\gamma^5$. The  current and its divergence are
\begin{equation}
\begin{aligned}
J_\mu(\eta'_{h})&=\overline{\Psi}\gamma_\mu\gamma_{5}I_{3\times3}\Psi , \\
\partial_{\mu}J^\mu(\eta'_{h})&=\frac{\sqrt{6}g^{2}}{16 \pi^{2}} G_{a}^{\mu \nu} \tilde{G}_{a \mu \nu}+\frac{\sqrt{6}g'^{2}}{16 \pi^{2}} {G'}_{A}^{\mu \nu} \tilde{G'}_{A \mu \nu},
\end{aligned}
\end{equation}
The corresponding potential of the composite axion at low energy is
\begin{equation}
V= B_1f_\pi ^4\left( \dfrac{\sqrt{6}\eta'_{h}}{f_{\pi_h}} +\theta  \right) ^2+ B_2f_{\pi_h} ^4\left( \dfrac{\sqrt{6}\eta'_{h}}{f_{\pi_h}} +\theta' \right) ^2,
\end{equation}
At the minimal point of the potential, the cancellation equation of $CP$ phases is
\begin{equation}
\theta_{phy}^{\prime}= \theta_{phy}= \theta+\sqrt{6}\dfrac{\left\langle\eta_{h}^{\prime}\right\rangle}{f_{\pi_h}}=0 .
\end{equation}

\bibliography{axionNote}

\providecommand{\href}[2]{#2}\begingroup\raggedright\begin{thebibliography}{10}

\bibitem{Hooft1976}
G.~{`}t Hooft, ``Symmetry breaking through bell-jackiw anomalies,''
  \href{http://dx.doi.org/10.1103/physrevlett.37.8}{{\em Physical Review
  Letters} {\bfseries 37} no.~1, (Jul, 1976) 8--11}.

\bibitem{Hooft1976a}
G.~{`}t Hooft, ``Computation of the quantum effects due to a four-dimensional
  pseudoparticle,'' \href{http://dx.doi.org/10.1103/physrevd.14.3432}{{\em
  Physical Review D} {\bfseries 14} no.~12, (Dec, 1976) 3432--3450}.

\bibitem{Baker2006}
C.~A. Baker, D.~D. Doyle, P.~Geltenbort, K.~Green, M.~G.~D. van~der Grinten,
  P.~G. Harris, P.~Iaydjiev, S.~N. Ivanov, D.~J.~R. May, J.~M. Pendlebury,
  J.~D. Richardson, D.~Shiers, and K.~F. Smith, ``Improved experimental limit
  on the electric dipole moment of the neutron,''
  \href{http://dx.doi.org/10.1103/physrevlett.97.131801}{{\em Physical Review
  Letters} {\bfseries 97} no.~13, (Sep, 2006) }.

\bibitem{Nelson1984NWCV}
A.~Nelson, ``Naturally weak {CP} violation,''
  \href{http://dx.doi.org/10.1016/0370-2693(84)92025-2}{{\em Physics Letters B}
  {\bfseries 136} no.~5-6, (Mar, 1984) 387--391}.

\bibitem{Barr1984StSwtPQS}
S.~M. Barr, ``Solving the strong {CP} problem without the peccei-quinn
  symmetry,'' \href{http://dx.doi.org/10.1103/physrevlett.53.329}{{\em Physical
  Review Letters} {\bfseries 53} no.~4, (Jul, 1984) 329--332}.

\bibitem{Gasser1982}
J.~Gasser and H.~Leutwyler, ``Quark masses,''
  \href{http://dx.doi.org/10.1016/0370-1573(82)90035-7}{{\em Physics Reports}
  {\bfseries 87} no.~3, (Jul, 1982) 77--169}.

\bibitem{Peccei1977}
R.~D. Peccei and H.~R. Quinn, ``{CP} conservation in the presence of
  pseudoparticles,'' \href{http://dx.doi.org/10.1103/physrevlett.38.1440}{{\em
  Physical Review Letters} {\bfseries 38} no.~25, (Jun, 1977) 1440--1443}.

\bibitem{Peccei1977a}
R.~D. Peccei and H.~R. Quinn, ``Constraints imposed by {CP} conservation in the
  presence of pseudoparticles,''
  \href{http://dx.doi.org/10.1103/physrevd.16.1791}{{\em Physical Review D}
  {\bfseries 16} no.~6, (Sep, 1977) 1791--1797}.

\bibitem{Weinberg1978ANLB}
S.~Weinberg, ``A new light boson?,''
  \href{http://dx.doi.org/10.1103/physrevlett.40.223}{{\em Physical Review
  Letters} {\bfseries 40} no.~4, (Jan, 1978) 223--226}.

\bibitem{Wilczek1978PoSitPoI}
F.~Wilczek, ``Problem of strong {P} and {T} invariance in the presence of
  instantons,'' \href{http://dx.doi.org/10.1103/physrevlett.40.279}{{\em
  Physical Review Letters} {\bfseries 40} no.~5, (Jan, 1978) 279--282}.

\bibitem{Bardeen1987}
W.~A. Bardeen, R.~Peccei, and T.~Yanagida, ``Constraints on variant axion
  models,'' \href{http://dx.doi.org/10.1016/0550-3213(87)90003-4}{{\em Nuclear
  Physics B} {\bfseries 279} no.~3-4, (Jan, 1987) 401--428}.

\bibitem{Kim1979a}
J.~E. Kim, ``Weak-interaction singlet and strong {CP} invariance,''
  \href{http://dx.doi.org/10.1103/physrevlett.43.103}{{\em Physical Review
  Letters} {\bfseries 43} no.~2, (Jul, 1979) 103--107}.

\bibitem{Shifman1980a}
M.~Shifman, A.~Vainshtein, and V.~Zakharov, ``Can confinement ensure natural
  {CP} invariance of strong interactions?,''
  \href{http://dx.doi.org/10.1016/0550-3213(80)90209-6}{{\em Nuclear Physics B}
  {\bfseries 166} no.~3, (Apr, 1980) 493--506}.

\bibitem{Dine1981a}
M.~Dine, W.~Fischler, and M.~Srednicki, ``A simple solution to the strong {CP}
  problem with a harmless axion,''
  \href{http://dx.doi.org/10.1016/0370-2693(81)90590-6}{{\em Physics Letters B}
  {\bfseries 104} no.~3, (Aug, 1981) 199--202}.

\bibitem{zhitnitskii1980possible}
A.~Zhitnitskii, ``Possible suppression of axion-hadron interactions,'' {\em
  Sov. J. Nucl. Phys.(Engl. Transl.);(United States)} {\bfseries 31} no.~2,
  (1980) .

\bibitem{RaffeltAAB}
G.~G. Raffelt,
  \href{http://dx.doi.org/10.1007/978-3-540-73518-2_3}{``Astrophysical axion
  bounds,''} in {\em Lecture Notes in Physics}, pp.~51--71.
\newblock Springer Berlin Heidelberg.

\bibitem{Barr1992}
S.~M. Barr and D.~Seckel, ``Planck-scale corrections to axion models,''
  \href{http://dx.doi.org/10.1103/physrevd.46.539}{{\em Physical Review D}
  {\bfseries 46} no.~2, (Jul, 1992) 539--549}.

\bibitem{Kamionkowski1992}
M.~Kamionkowski and J.~March-Russell, ``Planck-scale physics and the
  peccei-quinn mechanism,''
  \href{http://dx.doi.org/10.1016/0370-2693(92)90492-m}{{\em Physics Letters B}
  {\bfseries 282} no.~1-2, (May, 1992) 137--141}.

\bibitem{Hook2018TLotSCPaA}
A.~Hook, ``Tasi lectures on the strong {CP} problem and axions,''
  \href{http://arxiv.org/abs/http://arxiv.org/abs/1812.02669v1}{{\ttfamily
  http://arxiv.org/abs/1812.02669v1}}.

\bibitem{Kim2005}
J.~E. Kim, H.~P. Nilles, and M.~Peloso, ``Completing natural inflation,''
  \href{http://dx.doi.org/10.1088/1475-7516/2005/01/005}{{\em Journal of
  Cosmology and Astroparticle Physics} {\bfseries 2005} no.~01, (Jan, 2005)
  005--005}.

\bibitem{Choi2014a}
K.~Choi, H.~Kim, and S.~Yun, ``Natural inflation with multiple sub-planckian
  axions,'' \href{http://arxiv.org/abs/1404.6209v4}{{\ttfamily 1404.6209v4}}.

\bibitem{Higaki2016}
T.~Higaki, K.~S. Jeong, N.~Kitajima, and F.~Takahashi, ``Quality of the
  {Peccei-Quinn} symmetry in the aligned {QCD} axion and cosmological
  implications,'' \href{http://dx.doi.org/10.1007/JHEP06(2016)150}{{\em Journal
  of High Energy Physics} {\bfseries 2016} no.~6, (Jun, 2016) },
  \href{http://arxiv.org/abs/http://arxiv.org/abs/1603.02090v2}{{\ttfamily
  http://arxiv.org/abs/1603.02090v2}}.

\bibitem{Bonnefoy2018}
Q.~Bonnefoy, E.~Dudas, and S.~Pokorski, ``Axions in a highly protected gauge
  symmetry model,''
  \href{http://dx.doi.org/10.1140/epjc/s10052-018-6528-z}{{\em The European
  Physical Journal C} {\bfseries 79} no.~1, (Jan, 2019) },
  \href{http://arxiv.org/abs/http://arxiv.org/abs/1804.01112v2}{{\ttfamily
  http://arxiv.org/abs/1804.01112v2}}.

\bibitem{Fukuda2015AMoVQA}
H.~Fukuda, K.~Harigaya, M.~Ibe, and T.~T. Yanagida, ``A model of visible qcd
  axion,'' \href{http://arxiv.org/abs/1504.06084v2}{{\ttfamily 1504.06084v2}}.

\bibitem{Randall1992CAMaPSP}
L.~Randall, ``Composite axion models and planck scale physics,''
  \href{http://dx.doi.org/10.1016/0370-2693(92)91928-3}{{\em Physics Letters B}
  {\bfseries 284} no.~1-2, (Jun, 1992) 77--80}.

\bibitem{Strassler2006EoaHVaHC}
M.~J. Strassler and K.~M. Zurek, ``Echoes of a hidden valley at hadron
  colliders,''
  \href{http://arxiv.org/abs/http://arxiv.org/abs/hep-ph/0604261v2}{{\ttfamily
  http://arxiv.org/abs/hep-ph/0604261v2}}.

\bibitem{Kilic2009VCatL}
C.~Kilic, T.~Okui, and R.~Sundrum, ``Vectorlike confinement at the lhc,''
  \href{http://arxiv.org/abs/http://arxiv.org/abs/0906.0577v4}{{\ttfamily
  http://arxiv.org/abs/0906.0577v4}}.

\bibitem{Chacko2006NEBfaMS}
Z.~Chacko, H.-S. Goh, and R.~Harnik, ``Natural electroweak breaking from a
  mirror symmetry,''
  \href{http://dx.doi.org/10.1103/physrevlett.96.231802}{{\em Physical Review
  Letters} {\bfseries 96} no.~23, (Jun, 2006) }.

\bibitem{Weinberg1989LHBETitNEDM}
S.~Weinberg, ``Larger higgs-boson-exchange terms in the neutron electric dipole
  moment,'' \href{http://dx.doi.org/10.1103/physrevlett.63.2333}{{\em Physical
  Review Letters} {\bfseries 63} no.~21, (Nov, 1989) 2333--2336}.

\bibitem{Pospelov2005EDMAPoNP}
M.~Pospelov and A.~Ritz, ``Electric dipole moments as probes of new physics,''
  \href{http://arxiv.org/abs/http://arxiv.org/abs/hep-ph/0504231v2}{{\ttfamily
  http://arxiv.org/abs/hep-ph/0504231v2}}.

\bibitem{Rubakov1997}
V.~A. Rubakov, ``Grand unification and heavy axion,''
  \href{http://dx.doi.org/10.1134/1.567390}{{\em Journal of Experimental and
  Theoretical Physics Letters} {\bfseries 65} no.~8, (Apr, 1997) 621--624},
  \href{http://arxiv.org/abs/hep-ph/9703409v2}{{\ttfamily hep-ph/9703409v2}}.

\bibitem{Berezhiani2001a}
Z.~Berezhiani, L.~Gianfagna, and M.~Giannotti, ``Strong {CP} problem and mirror
  world: the weinberg{\textendash}wilczek axion revisited,''
  \href{http://dx.doi.org/10.1016/s0370-2693(00)01392-7}{{\em Physics Letters
  B} {\bfseries 500} no.~3-4, (Feb, 2001) 286--296}.

\bibitem{Albaid2015SCaS}
A.~Albaid, M.~Dine, and P.~Draper, ``Strong cp and suz$_2$,''
  \href{http://dx.doi.org/10.1007/JHEP12(2015)046}{{\em Journal of High Energy
  Physics} no.~12, (Dec, 2015) 1--18},
  \href{http://arxiv.org/abs/1510.03392v2}{{\ttfamily 1510.03392v2}}.

\bibitem{Hook2014ASttSCP}
A.~Hook, ``Anomalous solutions to the strong {CP} problem,''
  \href{http://dx.doi.org/10.1103/PhysRevLett.114.141801}{{\em Physical Review
  Letters} {\bfseries 114} no.~14, (Apr, 2015) },
  \href{http://arxiv.org/abs/1411.3325v2}{{\ttfamily 1411.3325v2}}.

\bibitem{Chiang2016}
C.-W. Chiang, H.~Fukuda, M.~Ibe, and T.~T. Yanagida, ``750 {GeV} diphoton
  resonance in a visible heavy qcd axion model,''
  \href{http://arxiv.org/abs/1602.07909v3}{{\ttfamily 1602.07909v3}}.

\bibitem{Choi1985DA}
K.~Choi and J.~E. Kim, ``Dynamical axion,'' {\em Physical Review D} {\bfseries
  32} no.~7, (1985) 1828.

\bibitem{Gherghetta2016a}
T.~Gherghetta, N.~Nagata, and M.~Shifman, ``A visible qcd axion from an
  enlarged color group,'' \href{http://arxiv.org/abs/1604.01127v3}{{\ttfamily
  1604.01127v3}}.

\bibitem{Agrawal2017ETfPCotQA}
P.~Agrawal, J.~Fan, M.~Reece, and L.-T. Wang, ``Experimental targets for photon
  couplings of the qcd axion,''
  \href{http://arxiv.org/abs/1709.06085v2}{{\ttfamily 1709.06085v2}}.

\bibitem{Sikivie1982ADWatEU}
P.~Sikivie, ``Axions, domain walls, and the early universe,''
  \href{http://dx.doi.org/10.1103/physrevlett.48.1156}{{\em Physical Review
  Letters} {\bfseries 48} no.~17, (Apr, 1982) 1156--1159}.

\bibitem{Kappl2014a}
R.~Kappl, S.~Krippendorf, and H.~P. Nilles, ``Aligned natural inflation:
  Monodromies of two axions,''
  \href{http://dx.doi.org/10.1016/j.physletb.2014.08.045}{{\em Physics Letters
  B} {\bfseries 737} (Oct, 2014) 124--128},
  \href{http://arxiv.org/abs/1404.7127v3}{{\ttfamily 1404.7127v3}}.

\bibitem{Kuster2008A}
M.~Kuster, G.~Raffelt, and B.~Beltr{\'{a}}n, eds.,
  \href{http://dx.doi.org/10.1007/978-3-540-73518-2}{{\em Axions}}.
\newblock Springer Berlin Heidelberg, 2008.

\bibitem{Cortona2015}
G.~G. di~Cortona, E.~Hardy, J.~P. Vega, and G.~Villadoro, ``The {QCD} axion,
  precisely,'' \href{http://dx.doi.org/10.1007/JHEP01(2016)034}{{\em Journal of
  High Energy Physics} {\bfseries 2016} no.~1, (Jan, 2016) },
  \href{http://arxiv.org/abs/1511.02867v2}{{\ttfamily 1511.02867v2}}.

\end{thebibliography}\endgroup
\bibliographystyle{utphys}
	
\end{document}